\documentclass[aps,showpacs,superscriptaddress,amsfonts,prfluids,floatfix]{revtex4-2} 
\usepackage{times}
\usepackage{graphicx}
\usepackage{color}

\begin{document}

\title{Helical microstructures in molluscan biomineralization are a biological example of close packed helices that may form from a colloidal liquid crystal precursor in a twist--bend nematic phase}

\date{Version of \today} 

\author{Katarzyna Berent}
\affiliation{Academic Centre for Materials and Nanotechnology,  AGH University of Science and Technology, 30-055 Krak\'ow, Poland}
\author{Julyan H. E. Cartwright}
\affiliation{Instituto Andaluz de Ciencias de la Tierra, CSIC--Universidad de Granada, 18100 Armilla, Granada, Spain}
\affiliation{Instituto Carlos I de F\'{\i}sica Te\'orica y Computacional, Universidad de Granada, 18071 Granada, Spain}
\author{Antonio G. Checa}
\affiliation{Departamento de Estratigraf{\'{\i}}a y Paleontolog{\'{\i}}a, Universidad de Granada, 18071 Granada, Spain}
\affiliation{Instituto Andaluz de Ciencias de la Tierra, CSIC--Universidad de Granada, 18100 Armilla, Granada, Spain}
\author{Carlos Pimentel}
\affiliation{Univ.\ Grenoble Alpes, Univ.\ Savoie Mont Blanc, CNRS, IRD, Univ.\ Gustave Eiffel, ISTerre, 38000 Grenoble, France}
\affiliation{Instituto Andaluz de Ciencias de la Tierra, CSIC--Universidad de Granada, 18100 Armilla, Granada, Spain}
\author{Paula Ramos-Silva}
\affiliation{Plankton Diversity and Evolution, Naturalis Biodiversity Center, Darwinweg 2, 2333 CR Leiden, Netherlands}
\author{C. Ignacio Sainz-D{\'{\i}}az}
\affiliation{Instituto Andaluz de Ciencias de la Tierra, CSIC--Universidad de Granada, 18100 Armilla, Granada, Spain}

\begin{abstract}
We demonstrate that nature has produced a close-packed helical twisted filamentous material in the biomineralization of the mollusc.
In liquid crystals, twist--bend nematics have been predicted and observed.
We present and analyse evidence that the helical biomineral microstructure of mollusc shells may be formed from such a liquid-crystal precursor.
\end{abstract}

\maketitle

\section{Introduction}

The shells of a particular superfamily of gastropod molluscs, the Cavolinioidea, together with some in the sister group, the Limacinoidea,  are formed of biomineral microstructures not found in any other molluscs. The shell microstructure is composed of helical filaments that fill space. Similar packing of helical rods or filaments has led to an amount of work looking at the close-packing geometry of filamentous and columnar materials \cite{grason2015}.
It is thus of interest to examine here a natural example of this packing. 
Moreover, similar helical structures have been predicted \cite{meyer1976} and  subsequently observed  \cite{cestari2011,borshch2013,chen2013,vcopivc2013} in liquid crystals, where they are termed twist--bend nematics.
We put forward and analyse the hypothesis that these  helical structures seen  in mollusc biomineralization might  have their origins in twist--bend liquid-crystal precursors.

\section{Helical fibres in euthecosomatous pteropod biomineralization}

\begin{figure}[tbp]
\begin{center}
\includegraphics*[width=\textwidth,clip=true]{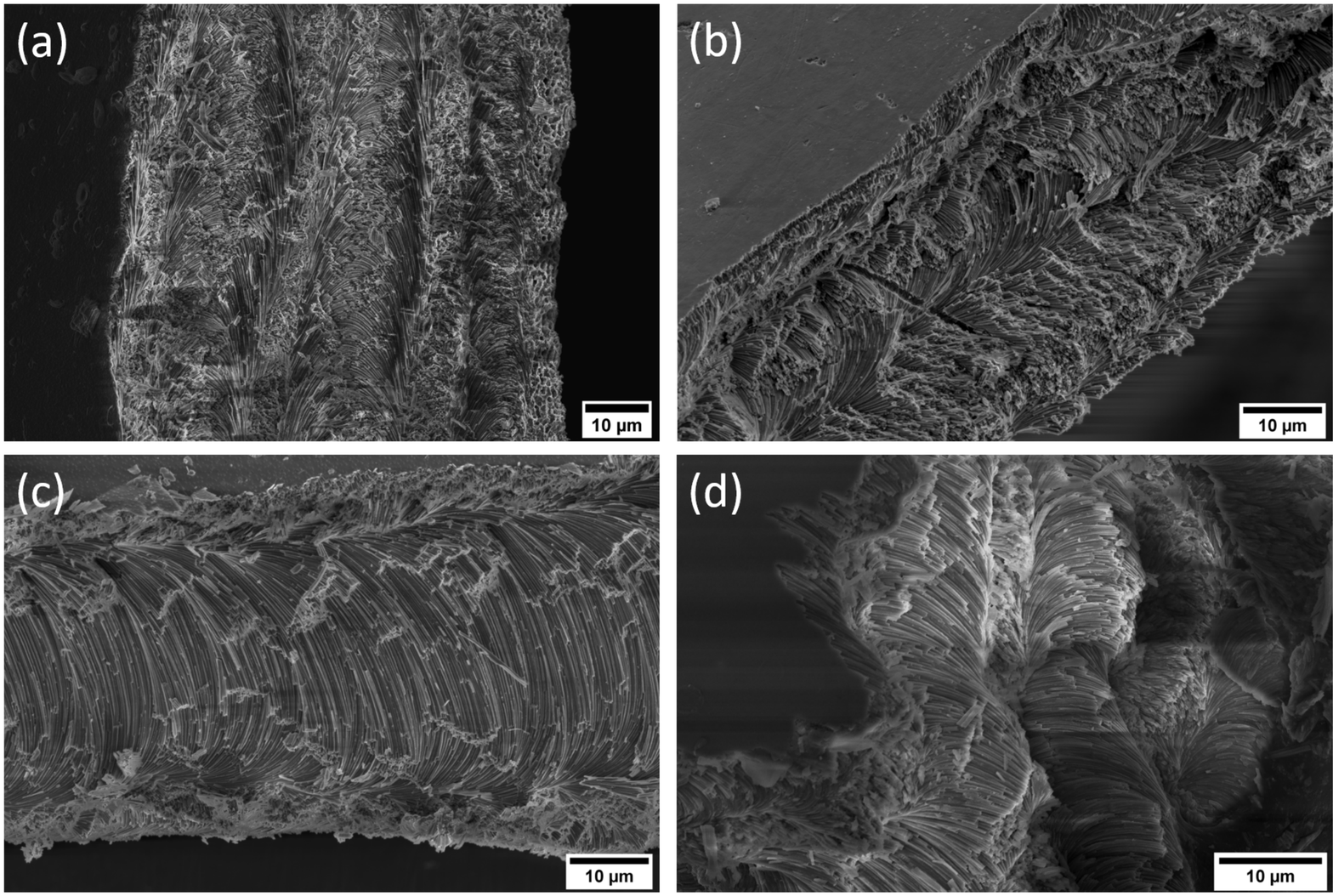}
\end{center}
\caption{\label{fig:fracture} 
Scanning electron micrographs of the aspect of the helical microstructure as seen in fracture surfaces in four species of cavolinioidean molluscs
(a) \emph{Diacria trispinosa},
(b) \emph{Clio sp},
(c) \emph{Cavolinia inflexa},
(d) \emph{Cuvierina columnella}.
}
\end{figure}

\begin{figure}[tbp]
\begin{center}
\includegraphics*[width=\textwidth,clip=true]{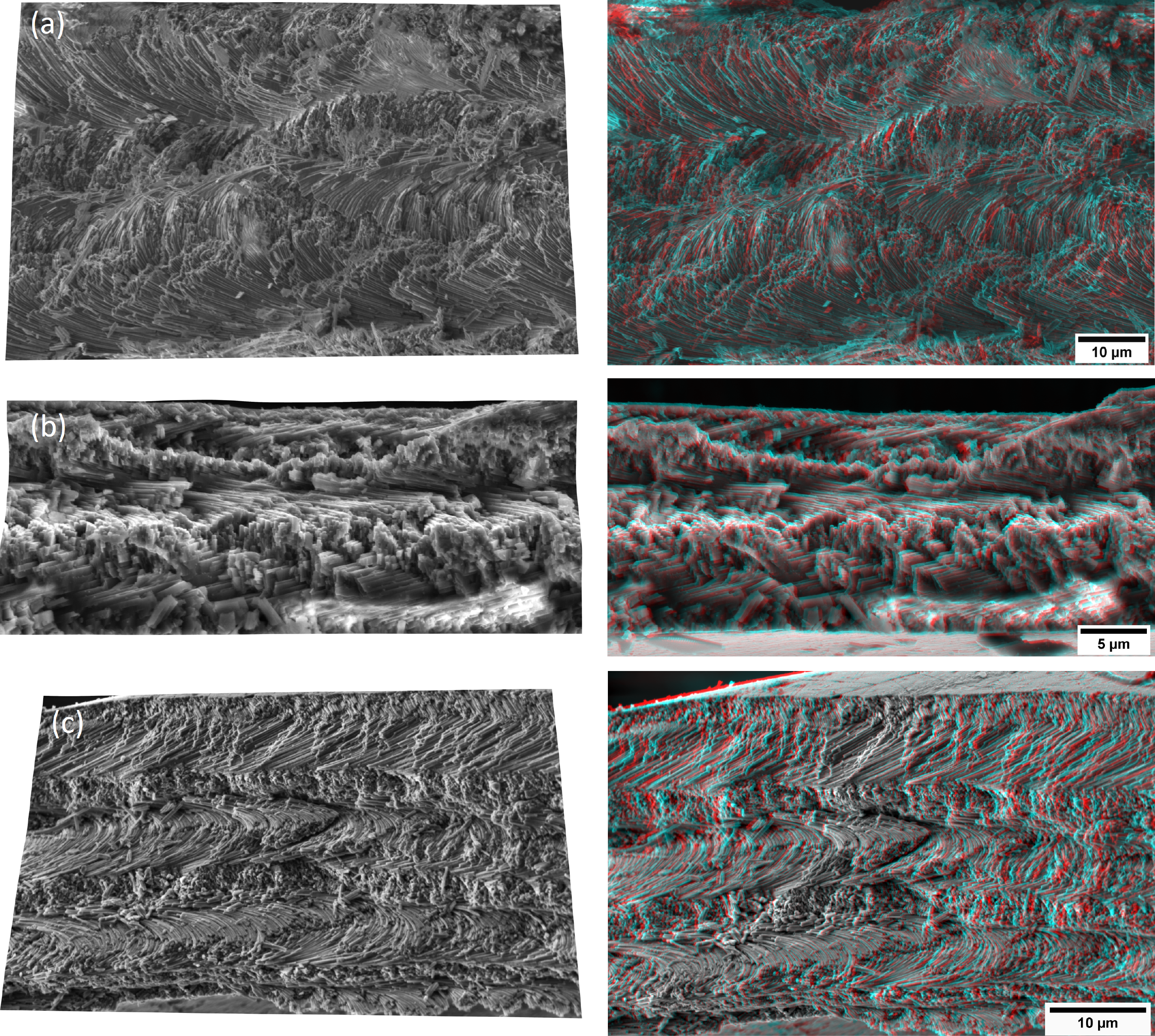}
\end{center}
\caption{\label{fig:mollusc3D} 3D reconstructions and anaglyph images from scanning electron micrography of three cavolinioidean mollusc species 
(a) \emph{Diacria trispinosa},
(b) \emph{Creseis acicula},
(c) \emph{Cavolinia inflexa}.
Further examples together with videos are provided in the SI.
}
\end{figure}

\begin{figure}[tbp]
\begin{center}
\includegraphics*[width=\textwidth,clip=true]{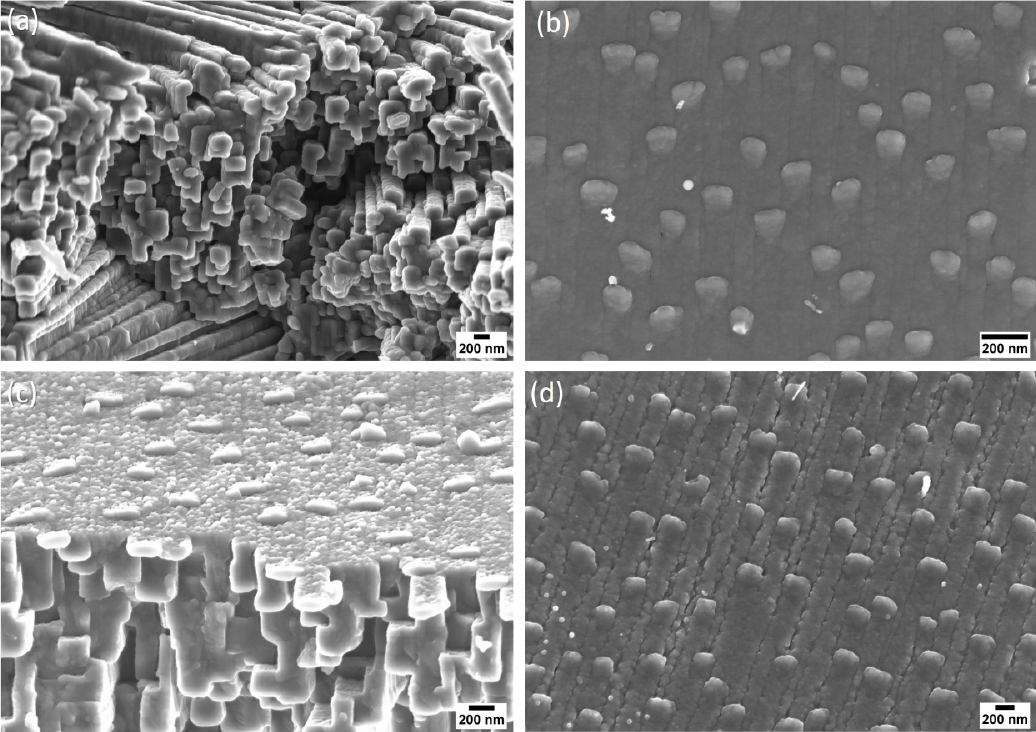}
\end{center}
\caption{\label{fig:growth} Scanning electron micrographs of the aspect of the helical microstructure as seen in growth surfaces 
in three species of cavolinioidean molluscs
(a,b) \emph{Clio sp},
(c) \emph{Cavolinia longirostris},
(d) \emph{Cuvierina columnella}.
}
\end{figure}

The aragonitic helical fibrous microstructure (helical microstructure from here on) is a most unusual molluscan microstructure (Fig.~\ref{fig:fracture}, \ref{fig:mollusc3D}). Its structure was first described in detail by B\'e et al. in 1972 \cite{be1972}. 
Recently, we  looked again at the microstructure  \cite{us2022}. 
It consists of very thin aragonite fibers which coil helically, in all cases dextrally when viewed from the shell outer surface,  from less than one turn to between three and four turns \cite{glacon1994,checa2016,ramossilva2021}. All fibers turn around axes parallel to each other and perpendicular to the shell surface and have similar helix parameters (radius, lead angle). Each fiber has its own coiling axis, which is slightly displaced with respect to those of its neighbors. During growth, all helical trajectories are in phase, i.e., at a given growth increment, they display exactly the same orientation. Fibers have constant widths, estimated at $\sim300$~nm, but they can extend in the shell-thickening direction --- toward the interior --- for very varied dimensions. This is because they have a complex interlocking mechanism \cite{zhang2011,li2015,checa2016}. In a section along the shell thickness, every fiber consists of two parts, a head and a tail. The head is the part (some 100--150~nm) looking toward the outer shell surface and the tail is the rest of the fiber, extending toward the shell interior. While the head is permanent (i.e., it is never intruded by a neighbor fiber during rotation), the tail is permanently changing its outline due to the interaction with the neighboring fibers. When we look at the shell growth surface, the heads appear as small protuberances, several tens of nanometers high (Fig.~\ref{fig:growth}a, c), while the tail is at the ground level (Fig.~\ref{fig:growth}b, d) \cite{checa2016}. 
If we consider a growth surface, everything corresponds to the same growth moment: the head and tail are coetaneous. On that plane, in the growth direction of the fiber, the head is the most forward position.
Interestingly, each fiber shows crystallographic continuity, and is composed of a myriad of aragonite crystals twinned on \{110\} \cite{willinger2016}. 
The degree of packing of fibers is total, such that they fill all the available volume. 

Although the helical microstructure was first described a half century ago, it has re-attracted the attention of materials scientists in the last two decades, not only because of its unusual organization and crystallography, but also due to its biomechanical performance. It has been found that this structure limits mechanical damage by creating tortuous crack propagation \cite{zhang2011,li2015}, thus making these ultra-thin shells an effective protection to the living tissues of the animal. 

To date, the helical microstructure has only been recognized in pteropods of the suborder Euthecosomata. This is a group of planktonic gastropods that swim in the plankton with extensions of the foot (parapodia). In relation to their planktonic habits, the shells are very thin and transparent. The suborder Euthecosomata is composed of two sister superfamilies: Limacinoidea and Cavolinioidea. The shells of limacinoideans are regularly coiled whereas those of cavolinioideans have variously-shaped non-coiled shells: straight or curved cone-, vase- or more complexly-shaped \cite{rampal1975,lalli1989}. This microstructure has been recognized in all members of the Cavolinioidea \cite{ramossilva2021} (composed of 63 species, 10 genera and 5 families; WoRMS Editorial Board 2022 \cite{worms2022}). It was also recognized in the limacinoidean genus \emph{Heliconoides}. 
In that genus, there is a crossed-lamellar layer at external positions \cite{ramossilva2021}. The crossed lamellar microstructure is the main component of the shells of the Limacinoidea. The concurrence of both microstructures suggests some evolutionary link between them.

\section{Packing helical filaments}

\begin{figure}[tbp]
\begin{center}
\includegraphics*[width=\textwidth,clip=true]{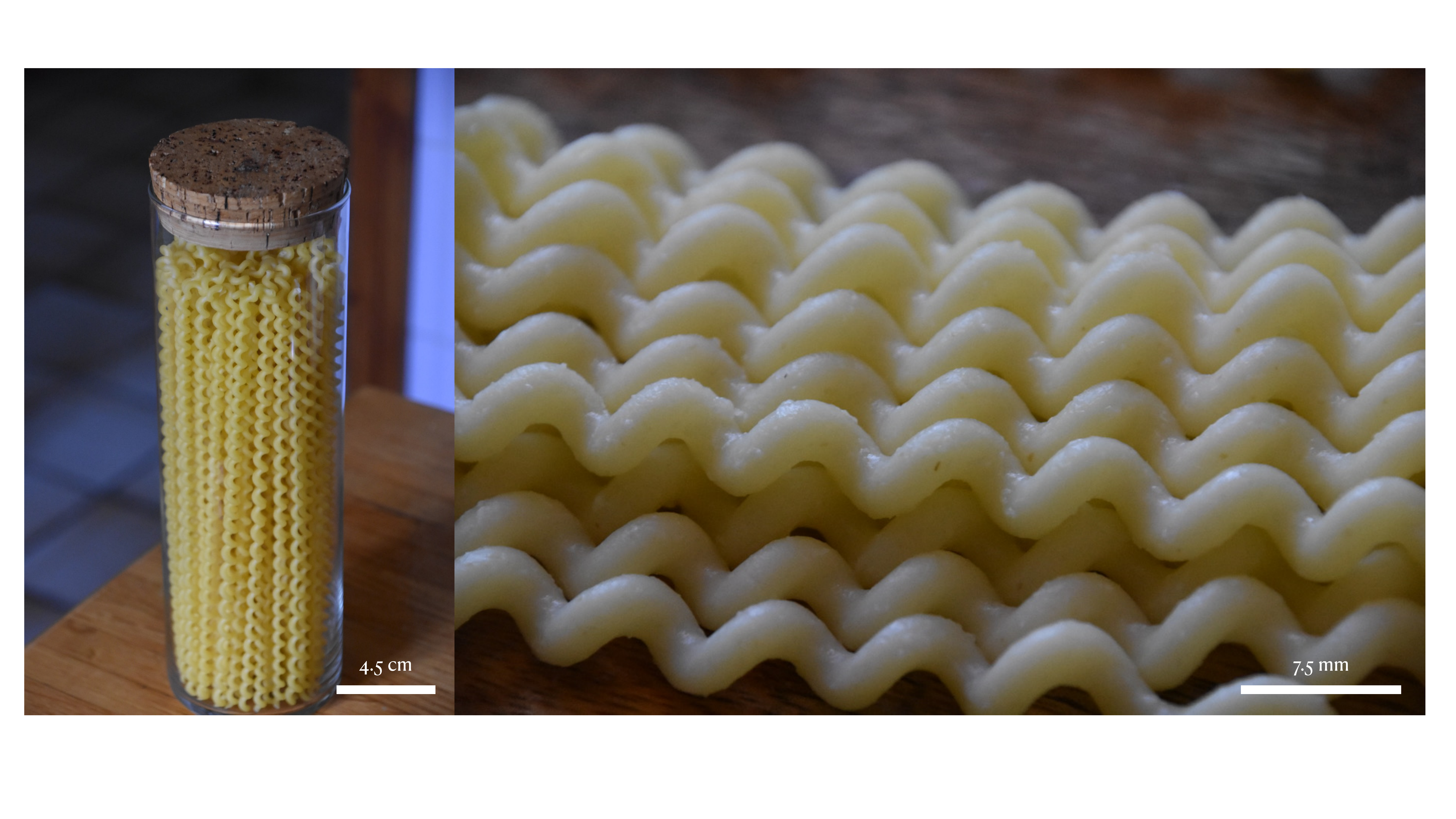}
\end{center}
\caption{\label{fig:pasta}
Packing twisted filaments. Helicoidal pasta: fusilli lunghi (left) in the uncooked, unpacked state and (right) cooked to become flexible and stretchable so that it may be packed  in a space filling array. The radius of the container is 4.5~cm and the wavelength of the pasta, 7.5~mm.
}
\end{figure}

\begin{figure}[tbp]
\begin{center}
\includegraphics*[width=0.9\textwidth,clip=true]{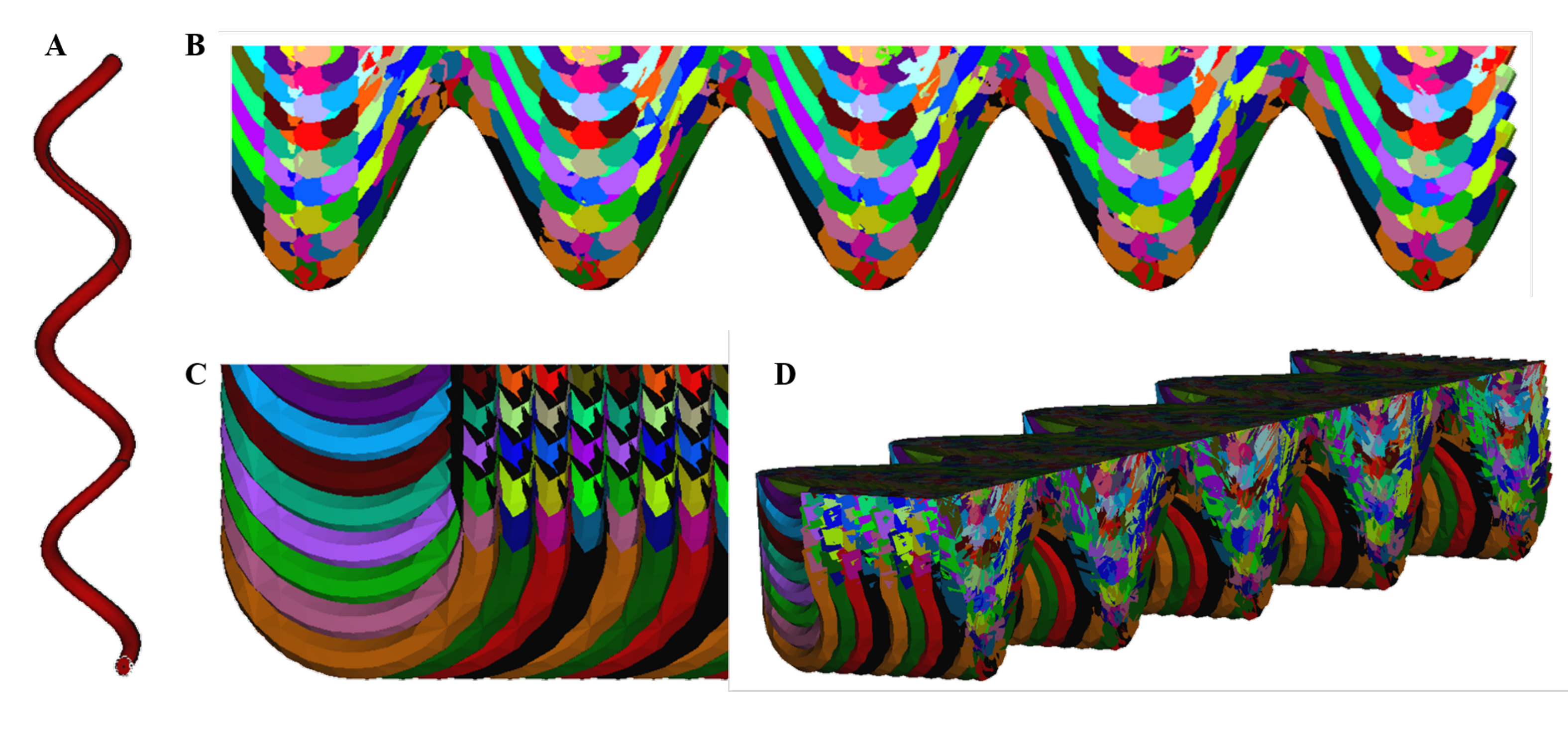}
\end{center}
\caption{\label{fig:isometric}
Isometric (constant spacing) packing.
(a) Model of a helix. (b--d) Packing of helical fibres cut along the axis of the helix along two orthogonal planes. All fibres are in full contact with each other, demonstrating the deformation produced in an isometric packing. The colours of the different fibres are to facilitate the identification of the individual fibres.
}
\end{figure}

\begin{figure}[tbp]
\begin{center}
\includegraphics*[width=0.9\textwidth,clip=true]{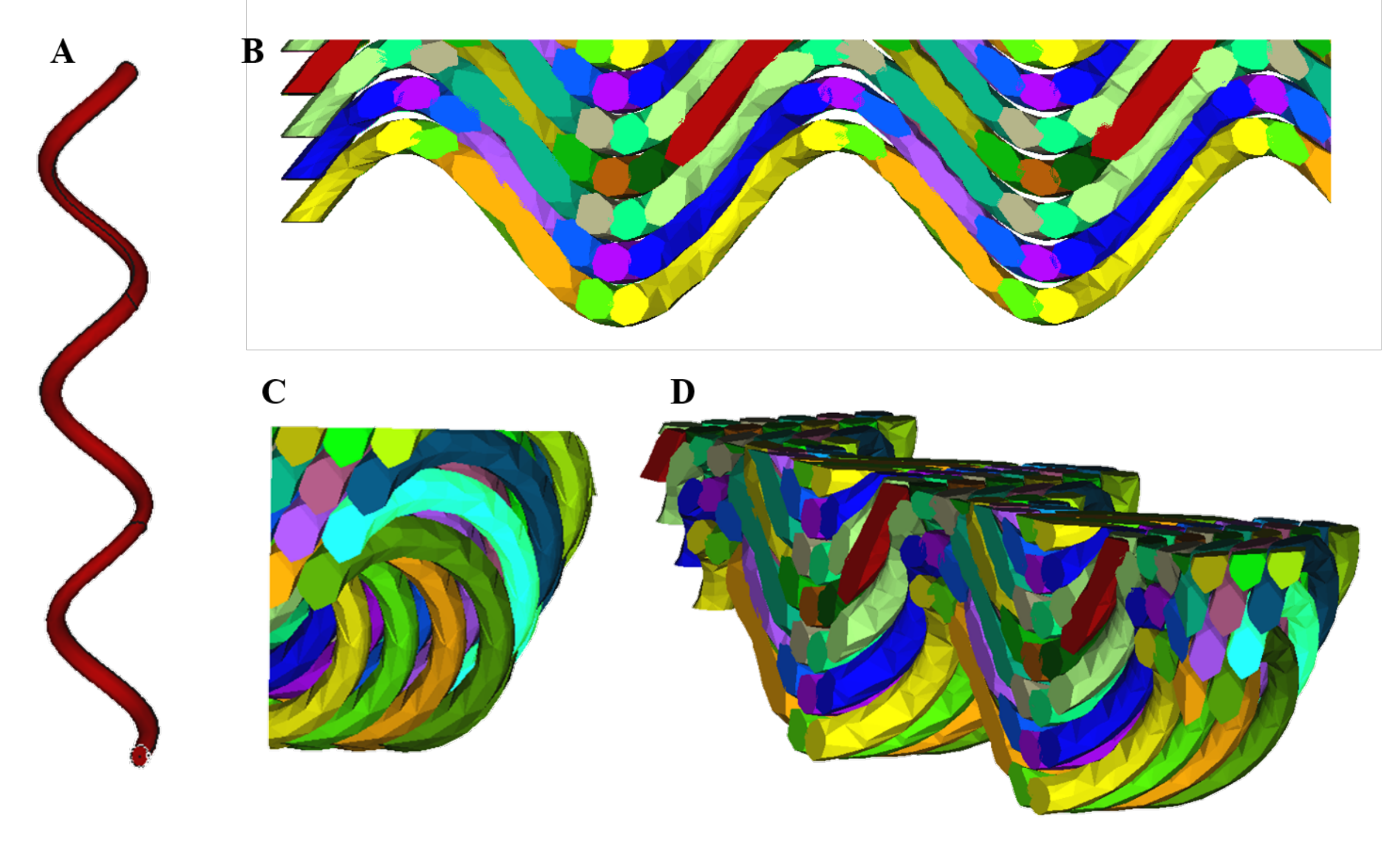}
\end{center}
\caption{\label{fig:isomorphic}
Isomorphic (constant shape) packing.
(a) Model of a helix. (b--d) Packing of helical fibres cut along the axis of the helix along two orthogonal planes. The helices are slightly in contact with each other, spaced such that identical filaments do not overlap in the bundle interior in an isomorphic packing. The colours of the different fibres are to facilitate the identification of the individual fibres.
}
\end{figure}

It is possible to grasp how helicoidal rods, packed in a hexagonal or a square array, may fill space by performing a thought experiment: imagine flexible, stretchable rods like cooked strands of spaghetti, initially all straight and packed together in such a fashion into a bundle of cooked pasta. Then, think of the whole bundle being bent and twisted at the same time so that along a filament the 2D planform translates in a circle. If the rods are flexible and stretchable they can be deformed in this fashion and a whole bundle of rods will stay together while being deformed. In fact, pasta makers have done the job for us. Helicoidal pasta exists, and we can fill space with fusilli lunghi cooked to become sufficiently flexible and stretchable, as we show in Fig.~\ref{fig:pasta}.
This flexibility and stretchability is needed, as noted in a  study \cite{grason2013} that distinguishes isometric (constant spacing) packing, Fig.~\ref{fig:isometric},  from isomorphic (constant shape) packing, illustrated  in Fig.~\ref{fig:isomorphic}. In fact,  a considerable body of theory  has been performed recently on the packing of twisted filaments  \cite{starostin2006,grason2009,olsen2010,grason2013,grason2015,panaitescu2018,atkinson2019}.  
While the packings discussed fill space, it may be noted that a different helicoidal packing results from rotating the 2D planform along a filament. The twisting of fibres around a central straight fibre in that configuration limits its radial growth \cite{weisel1987} so that to fill space one must group together smaller twisted elements into multiple strands or micelles \cite{pal2016}.  This latter type of packing is the basis of twisted cables and wire ropes \cite{bohr2011}, but is not what we see here.

It is  fascinating that here in pteropod mollusc biomineralization we have a  natural biological instance of  helicoidal space-filling filaments. By designing three-dimensional theoretical models, it is possible to study how both isometric and isomorphic packings behave when cut through different planes, which is the way in which the shells of these molluscs are studied. Thus, if a number of helices are grouped together to make a packing and cut along the axis of the helix using two orthogonal planes, it can be seen what these intersections look like in both isometric, Fig.~\ref{fig:isometric},  and isomorphic cases, Fig.~\ref{fig:isomorphic}.
 In these planes, two distinct zones can be observed, one zone composed of long fibre fragments and zones composed of multiple small fragments of different fibres. These distinct zones are clearly visible in the SEM images obtained when studying these shells, which allows us to deduce that these fibres are, in fact, helices.

\section{Liquid crystal precursors in biomineralization}

Many of the microstructures seen in the biominerals made by molluscs look similar to liquid-crystal structures. That similarity of form, plus the fact that molluscan biominerals form outside the cell, where they must necessarily self assemble,  has led us to investigate whether liquid crystal self-assembly is involved in the formation process.
A liquid crystal is a state of matter intermediate in some ways between a solid crystal and a liquid. In a colloidal liquid crystal, solid nanocrystals of a given size and shape, when present at sufficient concentration within a liquid, self organize within the liquid with a periodic pattern that minimizes their energy. A question that we are attempting to answer is: Are molluscs utilizing this physical mechanism of liquid-crystalline self-assembly to produce structures outside the cell?

More than a century ago, Haeckel wrote on liquid-crystalline order in biology \cite{haeckel1917}; Bouligand researched these issues for many years \cite{bouligand1968,bouligand1972}; and Neville's 1993 book \cite{neville1993} is full of examples of such pattern formation. Our research into this question has concentrated on the biomineralization processes in molluscs.
We began with nacre, the most well-known and researched molluscan biomineral microstructure, where there is now a significant body of evidence that this is so \cite{cartwright2007,cartwright2009}. The polysaccharide chitin is known to be involved in incipient nacre, where as chitin nanocrystals it forms the core of sheets, interlamellar membranes, that are laid down during nacre formation before mineral tablets fill the spaces between them. Chitin nanocrystals form colloidal liquid crystals under laboratory conditions when brought together at sufficient concentration \cite{belamie2004,tzoumaki2010,liu2018}. The evidence we have gathered suggests that a similar colloidal liquid crystal of chitin is the initial organizing structure around which nacre is formed.
Subsequently we looked at cuttlefish bone \cite{checa2015}, which possesses a so-called twisted plywood structure very similar to that seen in other organisms in which a cholesteric liquid-crystal precursor state may have played a part in the formation \cite{bouligand1972,neville1993}. 
In the crossed-lamellar microstructure, the currently available evidence is more sparse, but it also shows indications of having passed through a nematic liquid-crystal precursor state \cite{almagro2021}.
Here we put together the hypothesis that the helicoidal microstructure might have as a precursor a natural biological instance of a 
twist--bend nematic.

 \section{Twist--bend liquid crystals}
  
\begin{figure}[tb]
\begin{center}
\includegraphics*[width=\textwidth,clip=true]{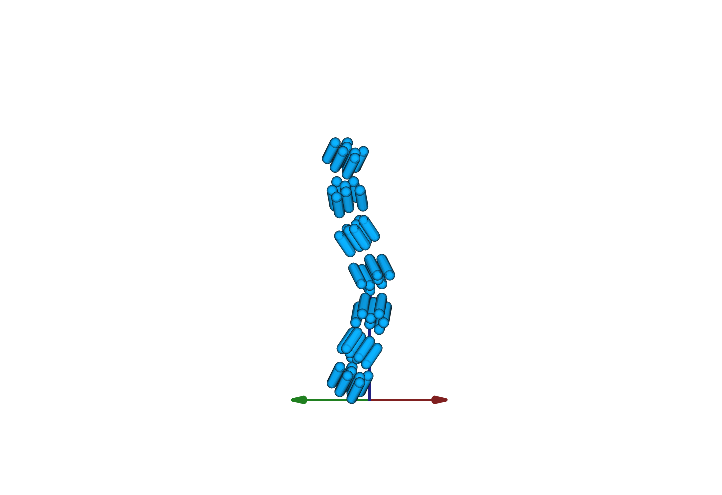}
\end{center}
\caption{\label{fig:twist-bend}
Twist--bend nematic.
Schematic representation of the twist--bend nematic liquid crystal structure.
}
\end{figure}

Although its existence had been predicted in the 1970s by Meyer \cite{meyer1976}, and also somewhat later in greater detail by Dozov \cite{dozov2001} and Memmer \cite{memmer2002}, until recently the twist--bend  liquid crystal phase (Fig.~\ref{fig:twist-bend}) had not been observed. In the last decade, however, twist--bend nematics have been reported in molecular liquid crystals \cite{cestari2011,borshch2013,chen2013,vcopivc2013}.  
It should be noted that there is controversy over whether these experimentally observed helicoidal structures should be termed twist--bend nematics or polar-twisted nematics \cite{samulski2020}. While we agree that this question needs to be addressed, 
to avoid confusion, here we are using the more commonly used former term.
Singularities and defects in twist--bend nematic liquid crystals have been investigated in a recent study \cite{binysh2020}.
There is now considerable excitement over potential technological  uses of the twist-bend nematic phase in electro-optical devices.

A chiral liquid-crystalline phase of helical filaments was reported by Barry et al.\ \cite{barry2006} in a colloidal liquid crystal formed from bacterial flagella. The difference there is that the elements of that liquid crystal, which they term a conical phase, are themselves helical, while here and in twist--bend  nematics in general the helical nature of the phase arises from the packing of elements, as shown in Fig.~\ref{fig:twist-bend}. As far as we know, the twist--bend phase has not as yet been reported in a colloidal liquid crystal. It is thus of particular interest to compare the helical microstructure of pteropod molluscs with the twist--bend  liquid crystal phase.   
   
The formation of the twist-bend nematic is associated with the
smallness of the bend elastic constant, which was indeed measured to
be very small as compared to other constants, see, e.g., Babakhanova et
al \cite{babakhanova2017}.  Since colloidal liquid crystals of chitin nanocrystals are not known to form twist-bend nematics on their own, we point to a chitin nanocrystal--protein complex as being a potential
source of bending flexibility.

\section{The formation of twist--bend mollusc biomineral microstructures}
  
 From what we have discussed in the preceding sections, together with what is known of the formation of other molluscan biomineral microstructures, 
 we  put forward a working hypothesis for how the helical microstructure found in mollusc biomineralization may form. 
 We suppose that first the cells of the mantle must expel into the liquid-filled extrapallial space some rod-shaped elements. This process is known from nacre biomineralization, where nanocrystals of the polysaccharide chitin initiate the formation of the interlamellar membranes that are the first structures to form in that microstructure \cite{cartwright2007,cartwright2009}. Chitin is widespread in molluscan biomineralization, and it is plausible that similar nanocrystals might be  present here. Similar processes of the formation of extracellular nanocrystals are known in the formation of that other common natural polysaccharide, cellulose, where rosette structures on plant cell walls directly crystallize cellulose from the monomer \cite{reis1991}. Once the chitin nanocrystals are present in sufficient concentration in the liquid, they should self-organize into a colloidal liquid crystal phase. Both nematic and cholesteric phases are known from laboratory experiments with chitin liquid crystallization \cite{belamie2004,tzoumaki2010}. Proteins are secreted to the extrapallial fluid by the mantle cells and occluded in the shell; 
  it is probable that it is the presence of a particular set of proteins at a given concentration that pushes the liquid crystallization into a given phase, in this case the twist--bend nematic that is in some sense intermediate between the nematic and cholesteric phases \cite{borshch2013}.
 
 In gastropod shells, there are chitin-binding proteins, but also several proteins with low-complexity regions including silk-like proteins and polyQ-rich proteins (glutamine-rich) with QQQQQQ (n-times) across their sequences, which are expressed in pteropods \cite{mann2014,ramossilva2021}. The Qn repetitive pattern  is known to form protein aggregates known as amyloid fibrils \cite{landrum2014} which have been shown to make colloidal gels with calcium carbonate \cite{shen2017}. This is evidence that not only chitin but also shell matrix proteins may form the colloidal liquid crystals, in the form of chitin-protein complexes.
 
 Once a liquid crystal phase has formed, it must act as the skeleton for biomineralization; it is a liquid-crystal precursor, because the final product is, of course, solid. In the case of nacre, where most is known, mineralization with calcium carbonate proceeds once the chitin--protein complex has been laid down \cite{cartwright2007}, and again, it is plausible that something similar is the case here: that after a twist--bend phase has formed and has filled the extrapallial space, calcium carbonate may grow around the chitin--protein skeleton, solidifying it.

\begin{figure}[tb]
\begin{center}
\includegraphics*[width=0.8\textwidth,clip=true]{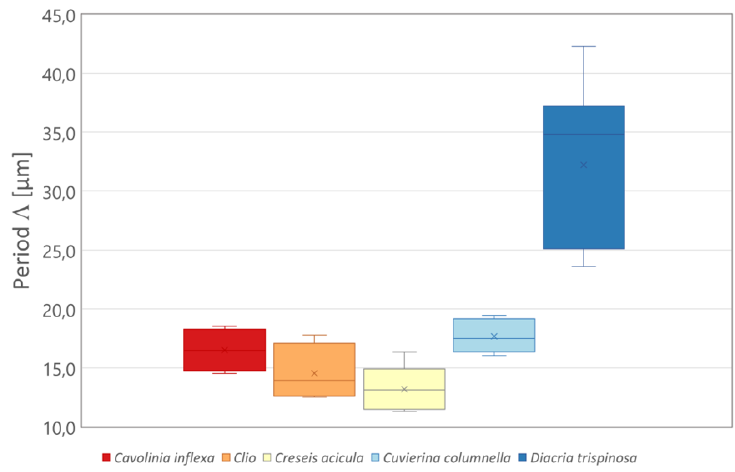}
\end{center}
\caption{\label{fig:wavelength}
Box plot of the period of the structure measured in 5 different species, \emph{Cavolinia inflexa}, \emph{Clio sp},  \emph{Creseis acicula},  \emph{Cuvierina columnella}, and \emph{Diacria trispinosa}.
}
\end{figure}

 In Fig.~\ref{fig:wavelength}  we present a quantitative analysis of the wavelength measured from SEM images. 
  The wavelength is different for different species, while there is also some variability between different specimens of the same species; in some species the variability is much less than in others. We may compare these micrometre wavelengths with the dimensions of chitin nanorods, some 20--30~nm in diameter.

 \section{Discussion}
 
Here, on the one hand we have seen that in pteropod mollusc biomineralization we have an instance of natural biological helicoidal space-filling filaments. On the other hand, we have put forward a hypothesis that these helicoidal space-filling filaments may form from a colloidal liquid crystal in the twist--bend nematic phase. The liquid crystals we are discussing we colloidal, rather than molecular liquid crystals. What we are probably dealing with in this complex biological system is a liquid crystal of chitin nanorods coated with protein. Are they straight or bent; do they possess chirality? We simply do not have data to answer all these questions at present. Instead in this natural system we are forced to work backwards, from the fact that we have the structures we show, to try to deduce the dynamical processes that led to their formation.
  
 The splay--bend nematic phase, composed of arc-shaped filaments, also predicted theoretically  
 has more recently been experimentally obtained with  banana-shaped colloidal elements \cite{fernandez2020}. 
It is possible that this phase might also be found in natural systems.
Curved fibres have also been described in the shell walls of other microgastropods including the holoplanktonic atlantid heteropods, known as sea elephants \cite{batten1976} or the scissurelids, known as little slit snails \cite{batten1975}. These microstructures were described as modified versions of the typical crossed lamellar  \cite{carter1990} being named type-2 crossed lamellar and were proposed to be a specific adaptation resulting from the thinness of the shell walls to provide optimal mechanical properties  \cite{batten1975}. But to be an arc-shaped microstructure we would need to have more than a curve in a single sense and the data currently available do not support the idea that these might be flat zig-zag curves.

\begin{figure}[tb]
\begin{center}
\includegraphics*[width=\textwidth,clip=true]{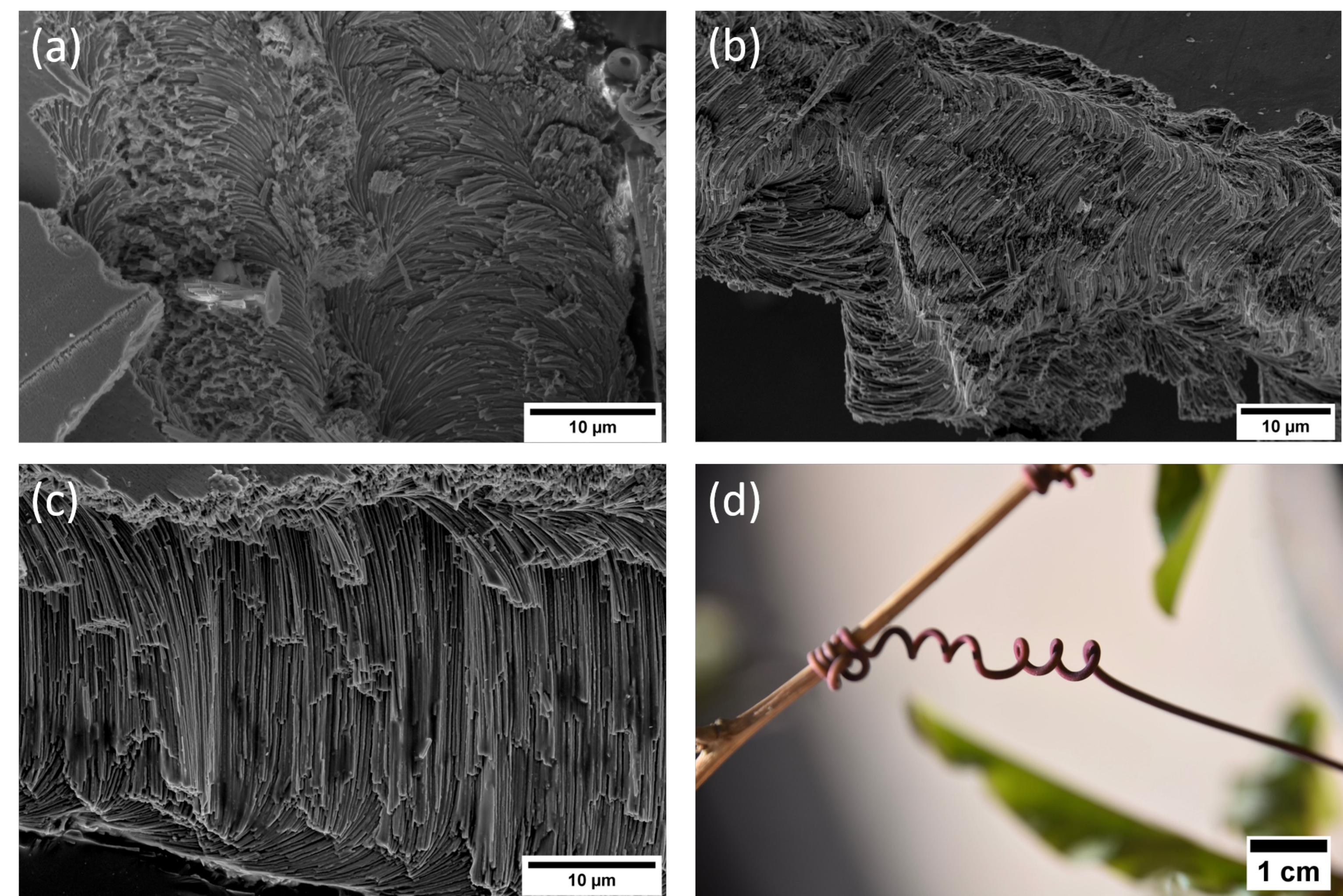}
\end{center}
\caption{\label{fig:Creseis}
Scanning electron micrographs of the aspect of the helical microstructure as seen in fracture surfaces in \emph{Creseis acicula}.
(a) from a juvenile  where the pattern seems to be purely helical,
(b) and (c) showing the  intermediate s-shaped stage unique to this species,
(d) an example of tendril perversion in \emph{Passiflora caerulea}.
}
\end{figure}

It has recently been proposed in molluscan biomineralization that the pteropod \emph{Creseis acicula} might have S-shaped arcs rather than helical shaped microstructure \cite{sibony2021}. 
Further imaging of \emph{Creseis acicula} shells, Fig.~\ref{fig:Creseis},  shows that in mature specimens the central section of the shell thickness in that species has a sinuous segment above and below which there are more normal helicoidal segments. They still appear 3D, rather than 2D: The curvature above and below the central section looks to be out of the plane of the central part.  
One possibility here is that we may be dealing with a topological defect, a biomineral instance of tendril perversion \cite{goriely1998,silva2016}, commonly observed in plants \cite{darwin1875} and in telephone cords. 
The intermediate S-shaped stage of such perversions which has a similar appearance to what we show in Fig.~\ref{fig:Creseis}
has been termed a hemihelix \cite{liu2014}. 
However, it is  difficult to be sure that we  have here a topological defect between left- and right-handed coiling; 
as Fig.~\ref{fig:Creseis} shows, 
there is not a long length of helix above and below this mid section. This feature, which appears to be unique among the species we have examined, would presumably be genetically determined in this species.

The first part of our work demonstrating the helicoidal structure can be added to other natural systems that show helicoidal space-filling filaments \cite{grason2015}, such as fibrillar collagen \cite{lillie1977,ruggeri1979,ottani2002} and fibrin bundles \cite{weisel1987,weisel2004},
and synthetic materials that have helical mesostructures \cite{yang2006}.
As for the second part, how can our hypothesis of a liquid crystal precursor to the solid biomineral be further tested? 
 How much chitin would need to be present in the shells with helical microstructures in order to propose it as a  scaffold-guide?
 Here we must note that chitin abundance is highly variable among molluscs. Some species have high amounts of chitin (e.g., 6.4\% of the total shell weight) \cite{goffinet1979,bezares2008}, whereas others, including cavolinioidean pteropods, have lower amounts of chitin when compared to the protein fractions in the shell \cite{poulicek1991,agbaje2018}. 
 One way of approaching this question is this: suppose that each filament grows around a single chitin nanocrystal --- or a chain of them, like a string of sausages --- at its core. From the SEM images, we see that a filament is some 200 to 300 nm in diameter. On the other hand, the nanocrystals of chitin plus protein observed in nacre are a tenth of this, 20--30 nm diameter. Since the cross sectional area goes as radius or diameter squared, that means the nanocrystal is a hundredth of the area. The volume proportion will be the same: 1\%, because these are filaments. This is the volumetric proportion. Considering that the density of the mineral is about 3 g/cm$^3$ and organic matter is similar to water, 1 g/cm$^3$, or less, the percentage weight should be 0.3\% or less.
This is all supposing that the nanocrystals are made up of the same number of chitin polymer chains as in nacre. If they 
have more or fewer polymer chains, or if they are not made of chitin, but of some other as yet undiscovered component, then things could be different. But at present the available evidence points to a chitin nanocrystal--protein complex being the probable originator of the liquid crystal precursor phase.

\section{Conclusions}

In the first part of this work, we have presented the  experimental evidence that the helical microstructure of the mollusc is  an instance from nature of close-packed helices that fill space, and we have performed modelling providing further evidence confirming that the shell structures studied are helices. In the second part, we put forward the  hypothesis   that this structure is, like some others found in nature, and in particular in molluscs, formed via a liquid-crystal twist--bend precursor phase that is later mineralized. Like all hypotheses, this one will stand or fall as further evidence is collected. That will require much careful interdisciplinary work on the organisms that produce this fascinating material.

\section{Methods}

\subsection{Scanning electron microscopy}
Microstructural studies of freshly broken specimens were performed using the field emission scanning electron microscopes (FESEM): Zeiss Auriga of the Center for Scientific Instrumentation (CIC) of the University of Granada (UGR), Spain and a FEI Versa 3D of the Academic Centre for Materials and Nanotechnology (ACMiN) of the AGH University of Science and Technology, Poland. The samples were cleaned with commercial bleach containing $\sim5$\% active chlorine for 4--5 min, then in distilled water, dried and coated with a nm layer of conductive carbon (Emitech K975X carbon evaporator) or directly observed using the low vacuum (LV) mode at 60~Pa chamber pressure.

\subsection{SEM wavelength meaurements}
We used 2 samples each of \emph{Diacria}, \emph{Clio}, and \emph{Creseis}, and 1 sample each of \emph{Cuvierina} and \emph{Cavolinia}.  We fragmented them and prepared them for imaging on SEM stubs. We chose 7 to 15 different fracture sections for the  measurements. From the thicker shells, \emph{Cuvierina} and \emph{Diacria} we took 15 different SEM images. For \emph{Creseis}, 14, for \emph{Clio}, 12, and 7 for \emph{Cavolinia}, which was the thinnest shell and which broke badly.  A summary of maximum and minimum values, average value and standard deviation is given in Table~\ref{table:wavelength}.

\begin{table}[tb]
\begin{center}
\caption{\label{table:wavelength}Table of maximum and minimum values, mean and standard deviation for the wavelength of helical structures in \emph{Cavolinia inflexa}, \emph{Clio sp},  \emph{Creseis acicula},  \emph{Cuvierina columnella}, and \emph{Diacria trispinosa}.}
\begin{tabular}{cccccc}
\hline
Period [$\mu$m] &	\emph{Cavolinia inflexa}	& \emph{Clio}	& \emph{Creseis acicula}	& \emph{Cuvierina columnella}	& \emph{Diacria trispinosa} \\
\hline
Minimum	&14.52 &12.52 &	11.31	&16.04&	23.60 \\
Maximum	&18.55 &	17.79	&16.34&	19.46&	42.27 \\
Mean &	16.51 &	14.55&	13.18&	17.67&	32.21 \\
Standard deviation &	1.60 &	2.09 &	1.78	& 1.25&	6.72 \\
\hline
\end{tabular}
\end{center}
\end{table}

\subsection{3D images}
Scanning electron microscope (SEM) images were acquired at a tilt angle of 0$^{\circ}$, then tilted along a vertical axis at 5$^{\circ}$ and/or 8$^{\circ}$ using a FEI Versa 3D (ACMiN, AGH UST) field emission scanning electron microscope (FESEM) operating at 5 or 10~kV. The pairs of SEM images were merged to make a red-cyan anaglyph image and 3D projection of the shell surface using MountainsLab 9 software (Digital Surf). A number of these 3D projections were taken with different inclinations, which were afterwards processed using the 3DF Zephyr Lite 6.5 trial version (3DFlow) software. Using 3DF Zephyr, 3D models of the surfaces were created. See the Supplemental Material \cite{SI} for more examples. The videos shown in the Supplemental Material were also recorded using this software. 3D reconstructions were constructed by using the following presents: for camera orientation, urban and deep; for dense point cloud creation and surface reconstruction, urban and high details; and for texturing, general and high details.

\subsection{3D helix models}
The design of the 3D models of the helix was carried out using FreeCAD software, the free and open source version of AutoCAD. FreeCAD allows us to design 3D models of a helix and a set of helices with different packaging arrangements. Such models allow us to describe the packaging and the intersection of the helices with different planes. The model was designed in the Part workbench using the Helix tool. The radius of the spiral was set at 10~mm, the height at 150~mm and the pitch at 30~mm. To give volume to the helix, we used the sketch tool in the Design Part workbench. The work plane chosen was XZ plane and a circle was drawn. By using the sweep tool in the Part workbench, the volume was created using the circle as the profile, the helix as the trajectory and choosing the solid and frenet options. Afterwards, the helix bulk models were constructed by replicating these initial helices. In one of the models the helices could interpenetrate, in an attempt to simulate isometric packing, while in the other model, the helices could not interpenetrate, to simulate isomorphic packing. The models were cut with orthogonal planes to allow the visualization of the intersections between these planes and the designed 3D models.

\subsection{Helical pasta packing experiments}
Approximately 3~L of water was placed in a large pot, $\sim 0.5$~g of sodium chloride was added and the water was heated to boiling.  Helical pasta (500~g Fusilli Lunghi Bucati di Napoli,  Terre d'Italia)
was added to the water,  stirring to submerge it and to prevent it from sticking to itself and to the container. The pasta was strained from the water after a given length of time and used for the packing experiment as soon as it was cool enough to handle (a few seconds): Pasta cooked `al dente' for eating ($\sim 540$~s) was suitable for isomorphic packing experiments; for isometric packing it was found to be necessary to boil it until almost mushy ($\gtrsim 660$~s).

\section{Acknowledgements}

CP has received funding from the Spanish Ministry of Science through Juan de la Cierva-Formaci\'on-2018 postdoctoral contract ref. FJC2018-035820-I and from the European Union Horizon 2020 research and innovation programme under the Marie Sklodowska-Curie grant agreement No. 101021894 [CARS-CO2].
PRS has received funding from the European Union Horizon 2020 research and innovation programme under the Marie Sklodowska-Curie grant agreement No. 844345 [EPIC]. 
KB and AC were funded by the Polish National Agency for Academic Exchange, grant PPI/APM/2018/1/00049/U/001. JHEC would like to acknowledge the contribution of the COST Action CA17139, Eutopia.

\bibliographystyle{aipnum4-1}
\bibliography{CL}

\end{document}